\documentclass[twocolumn]{article}
\usepackage[top=1in, bottom=1in, left=1in, right=1in]{geometry}
\usepackage{graphicx} 
\usepackage{multirow}
\usepackage{makecell}
\usepackage[version=4]{mhchem}
\usepackage{pdflscape} 
\usepackage{afterpage} 
\usepackage{enumitem}

\title{The Combination of Metal Oxides as Oxide Layers for RRAM and Artificial Intelligence}
\author{Sun Hanyu\thanks{Email: hanyu.sun@polyu.edu.hk
} }
\date{March 2023}

\begin{document}

\maketitle

\begin{abstract}
Resistive random-access memory (RRAM) is a promising candidate for next-generation memory devices due to its high speed, low power consumption, and excellent scalability. Metal oxides are commonly used as the oxide layer in RRAM devices due to their high dielectric constant and stability. However, to further improve the performance of RRAM devices, recent research has focused on integrating artificial intelligence (AI). AI can be used to optimize the performance of RRAM devices, while RRAM can also power AI as a hardware accelerator and in neuromorphic computing.
This review paper provides an overview of the combination of metal oxides-based RRAM and AI, highlighting recent advances in these two directions. We discuss the use of AI to improve the performance of RRAM devices and the use of RRAM to power AI. Additionally, we address key challenges in the field and provide insights into future research directions.
\end{abstract}

\section{Introduction}
Resistive random-access memory (RRAM) has emerged as a promising technology for next-generation memory devices due to their distinct features like high speed, ultra-low power consumption, and small footprint. RRAMs can be fabricated in a three-layer metal-insulator-metal configuration that uses various materials, among which metal oxides as the switching layer owning the advantage of simple chemical composition, multistate switching ability, and CMOS compatibility for the fabrication of RRAM devices \cite{patil2023binary}. These advantages make metal oxides RRAM potential in breaking the performance bottleneck of traditional architectures and dramatically increase computing power that is suitable for logic computing, neural networks, brain-like computing, and fused technology  of  sense-storage-computing \cite{wang2023research}. However, challenges still exist in terms of improving device performance and reliability, such as addressing issues related to endurance, retention, and variability.

Given the above, research has been conducted to study the characteristics of metal oxides RRAM, early in 2012, Wong \cite{wong2012metal} studied the physical mechanism, material properties, and electrical characteristics of a variety of binary metal–oxide RRAM with a focus on the use for non-volatile memory application. Besides mechanism and properties, Shi \cite{shi2021review} lately studied technologies that can guide device design and various applications based on resistive switching devices. Recently, Patil \cite{patil2023binary} made research about binary metal oxide-based resistive switching memory devices in which multiple aspects and factors responsible for the improved performance of memory devices have been deliberated critically. This existing research produces knowledge that is crucial in optimizing metal oxides-based resistive random-access memory (RRAM) and then achieving better memory performance. However, the existing knowledge base related to metal oxides-based RRAM devices is often diverse and heterogeneous, which can be overwhelming for researchers who are just starting their careers \cite{dongale2022machine}.

To further improve the performance of RRAM devices and meet the demands of future computing systems, artificial intelligence (AI) has been proposed as a novel approach. For example, machine learning (ML) can be used to provide guidelines to assist researchers in the optimization process by first analyzing large amounts of data from experiments, simulations, and literature to extract patterns and relationships between different variables that affect metal oxides-based RRAM performance, then using this information to guide the design and fabrication of high-performance metal oxide-based RRAM devices. Furthermore, ML techniques can also be used to optimize the fabrication process itself, by predicting the optimal fabrication parameters for metal oxides-based RRAM devices. This can lead to the faster and more efficient development of high-performance metal oxide-based RRAM devices. Meanwhile, the semiconductor manufacturing industry is also going through a phase of automation like Industry 4.0, a fourth industrial revolution, proposes the idea of intelligent manufacturing by implementing the Internet of things (IoT) and Artificial intelligence (AI) technologies for smart fabrication \cite{zhong2017intelligent}. For example, in the regime of process-aware compact device modeling, a single model encompasses the prediction from the process, material, and device parameters to device current-voltage (I-V) characteristics is needed but difficult to realize using conventional methods thus giving the opportunity to ML-based CM. Specific for metal oxides based RRAM fabrication, Lin and Hutchins \cite{lin2020process,lin2022rram,hutchins2022generalized} have taken actions.

With data-driven ML, continuous innovation in the area of metal-oxide RRAM is possible. In addition, RRAM devices can be used to power AI by improving the speed, accuracy, and energy efficiency of AI algorithms and making them suitable for a wide range of applications. Therefore, it is important to present metal oxides-based RRAM devices to an AI scientific community in a holistic manner. Given this, this review paper provides an overview of the combination of metal oxides as oxide layers for RRAM and AI, highlighting recent advances in these two directions.

\section{Metal oxides for RRAM}

\begin{figure}
  \centering
    \includegraphics[width=0.8\linewidth]{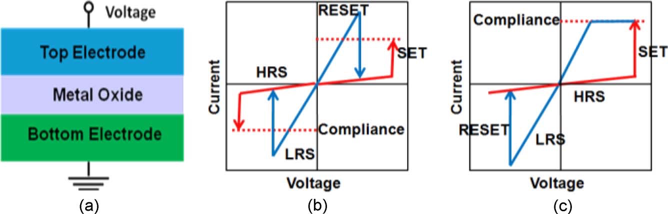}
  \caption{(a) Schematic of MIM structure for metal–oxide RRAM, and schematic of metal–oxide memory’s I–Vcurves, showing two modes of operation: (b) unipolar and (c) bipolar. \cite{wong2012metal}}
  \label{im1}
\end{figure}

The use of metal oxides as the oxide layer in RRAM devices has been widely explored. Typical structure of the metal oxides-based device is simply an oxide material sandwiched between two metal electrodes,called the metal–insulator–metal (MIM) structure as shown in Figure~\ref{im1}. Various types of metal oxides materials have been explored by the industry and academia for fabricating high-performance RS devices, among which, the binary metal-oxides including \ce{TiO2}, \ce{ZnO}, \ce{NiO}, \ce{WO3}, \ce{TaOx}, \ce{Ta2O5}, \ce{HfO2}, and \ce{CuOx} were extensively reported as a suitable active layer or switching material.

\begin{figure}
  \centering
    \includegraphics[width=0.8\linewidth]{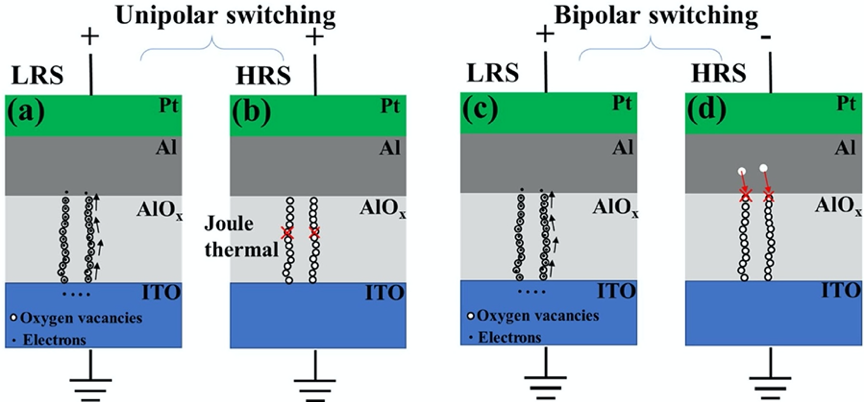}
  \caption{(a-d): A schematic illustration of the thermo-chemical mechanism of the AlOx-based RRAM device in unipolar and bipolar mode. \cite{patil2023binary}}
  \label{im2}
\end{figure}

Figure~\ref{im2} (a-d) represents a schematic illustration of the thermal-chemical mechanism (TCM) of the \ce{AlOx}-based RRAM device in unipolar and bipolar modes \cite{patil2023binary}. Under the action of the electric field, the oxygen ions in the switching layer move toward the top electrode(TE). After reducing these ions, only oxygen vacancies remain in the switching layer. The sufficient accumulation of these oxygen vacancies in the switching layer forms the oxygen-based CFs between the bottom electrode (BE) (indium tin oxide, ITO) and the unoxidized metal ions layer (Al). As a result, the device switched from the HRS to LRS (SET).

In the context of RS devices, in addition to types of materials, there are many other features which can be divided to continuous and categorical/discrete features. Continuous variables refer to physical or electrical properties that can take on any value within a certain range, such as resistance or conductance. These variables can be measured and quantified using numerical values.
On the other hand, categorical variables refer to properties that are divided into discrete categories or classes, such as the type of metal oxide used as the resistive layer in an RS device or the type of measurement equipment used to characterize the device. These variables are not numerical and cannot be measured on a continuous scale. 
Both continuous and categorical variables are important in characterizing RS devices and understanding their behavior. Continuous variables can provide quantitative measures of the device's performance, while categorical variables can help to identify trends and patterns in the data that may be associated with specific device properties or fabrication processes. 

A detailed collection of features used by papers is shown in Table~\ref{table1}. In the continuous features, 
the Vegard strain coefficient, electrical conductivity, thermal conductivity, and current on/off ratio are generally treated as a constant value for a given device.

\begin{table*}[]
\centering
\caption{Features of metal-oxides as oxide layers for RRAM}
\small
\begin{tabular}{|c|p{10cm}|}
\hline
\multirow{3}{*}{\makecell[c]{\\ \\ Type of RS material \\ (TYM)}} & Top electrode   (TE): \ce{Sn}, \ce{Si}, \ce{Hf}, \ce{Zr}, \ce{Cr}, \ce{IrOx}, \ce{In}, \ce{V}, \ce{Ru}, \ce{Co}, \ce{TiW}, \ce{Ir}, \ce{TaN}, \ce{Pd}, \ce{ITO}, \ce{W}, \ce{Al},   \ce{Ag}, \ce{Ni}, \ce{Au}, \ce{Ta}, \ce{Ti}, \ce{Cu}, \ce{TiN}, \ce{Pt}, et al.                                                                                                                                                                                                               \\
                                           & Bottom electrode (BE): \ce{WNx}, \ce{SS}, \ce{SiGe}, \ce{HfMx}, \ce{CuTe}, \ce{RuOx}, \ce{ZnO},   \ce{TiSi2}, \ce{CoSi2}, \ce{Cu}, \ce{AI}, \ce{TaN}, \ce{Ti}, \ce{CNT}, et al.                                                                                                                                                                                                                                          \\
                                           & Oxide layer: \ce{HfOx}, \ce{Ta2O5}, \ce{TaOx}, \ce{Al2O3}, \ce{SiO2}, \ce{NiOx}, \ce{ZnO}, \ce{WOx}, \ce{ZrOx}, et al.                                                                                                                                                                                                                                                                      \\ \hline
\multirow{1}{*}{\makecell[c]{\\ \\ Categorial features} } & Type of   materials/device structures, type of switching (analog or digital) (TSAD),   types of RS present in the RS devices (TSUB)(unipolar or bipolar), synthesis   method (SM), fabrication methods used to prepare active switching layer,   multiple resistance states (MRS), conduction mechanism (CM), resistive   switching mechanism (RSM). \\ \hline
\multirow{1}{*}{\makecell[c]{\\ \\  Continuous features } }         & Thickness   distribution of TE and BE (TTE, TBE), the thickness distribution of active   switching layer (TSL), switching voltage distribution, nonvolatile memory   performance of RS devices, SET voltage, RESET voltage, endurance, retention,   and memory window, Vegard strain coefficient ($V_{ij}$), electrical conductivity ($\sigma$), thermal conductivity $k_{th}$, current on/off ratio ($I_{on}/I_{off}$), resistance switching time ($t_{switch}$).                                                                              
\\ \hline
\end{tabular}
\label{table1}
\end{table*}

\section{AI for RRAM}

Though AI can be used to optimize the performance of RRAM devices, few researchers have tried it. Zhang \cite{zhang2020high} developed a machine-learning model using high-throughput phase-field simulation data to predict the performance of $\textit{HfO}_2$ based RRAM devices, then the built model was used to select materials for improving RRAM performance. Lin \cite{lin2020process} tried a long short-term memory (LSTM) based compact model to fit the current-voltage (I-V) characteristics for RRAM. His model provided an alternative way  in contrast to physics-based CMs and was thought to have the potential of competing with the conventional compact device models in terms of shorter developing time, and better fitting capability in emerging devices. Lately, Lin \cite{lin2022rram} proposed a physics architecture in MLCMs for RRAM. The results show that the physics-assisted architecture enables simpler ML models in reference to our previous work of LSTM-based RRAM CMs. Moreover, Hutchins \cite{hutchins2022generalized} proposed a data augmenting method that duplicated the data and added Gaussian noise to the data to improve the performance of the model. Recently, Dongale \cite{dongale2022machine} used ML to provide design guidelines and predicted the performance of industry-standard resistive switching (RS) memory devices based on several Metal Oxides materials, in his study, heterogeneous data from previous publications was used which is different from previous used experimental data or simulations. Moreover, Suvarna \cite{patil4303488unraveling} used a similar method to predict a different metal oxide's performance and achieved good results. 

Thought ML were used in the above studies, different purposes were realized. For example, Zhang, Tukaram and Suvarna mainly focused on the discovery and understanding of different RS properties then able to accelerate novel devices' development, while Lin and Hutchins were interested in the device models then predicting RS devices' properties. A detail comparison can be seen in Table~\ref{table2}.

\subsection{Dataset Selection}

There is a lack of open-source repositories/datasets for RS devices that could be used for ML investigation. To use ML to train and test models, first-principle calculations, density functional theory, and computational simulations are typically used to generate datasets. 
Unfortunately, the aforementioned computational techniques produce homogeneous datasets with limited real-life significance.
In  some  cases,  homogeneous  datsets  are  used by simulating various models, or by computing the density functional/first principle theories or utilizing lab datasets,  owing  to  the  easy availability of raw data. However, such datasets  cannot  provide  detailed  insights  into  the materials  or  devices,  owing  to  the  absence  of intrinsic randomness. The heterogeneous dataset which can be created by collecting the raw data manually/automatically from the peer-reviewed literature can be the best way to identify the  patterns  and  understand  the  hidden  dynamics  of  the  materials  or  devices \cite{patil4303488unraveling}.

\subsection{Machine Learning Models}

The ML models used above include MLP, CART, DT, and LSTM which are very basic and have been frequently used in many other areas.

\begin{itemize}[leftmargin=*]
\item \textbf{Multilayer Perceptron (MLP)} A MLP is a fully connected class of feedforward artificial neural network (ANN) which  consists of at least three layers of nodes: an input layer, a hidden layer, and an output layer. MLP utilizes a supervised learning technique called backpropagation for training.

\item \textbf{Decision Tree (DT)} Decision tree learning is a supervised learning approach used in machine learning. Tree models where the target variable can take a discrete set of values are called classification trees; Decision trees where the target variable can take continuous values are called regression trees. More generally, the concept of regression trees can be extended to any kind of object equipped with pairwise dissimilarities such as categorical sequences.

\item \textbf{Classification and Regression Trees (CART)} CART is a type of decision tree algorithm used in machine learning. CART algorithms are used for both classification and regression problems. CART can handle both categorical and continuous variables, and can automatically handle missing data by imputing values during the tree-building process.

\item \textbf{Long short-term memory (LSTM)} LSTM is a type of recurrent neural network (RNN) architecture that is designed to better handle the vanishing and exploding gradient problem, which is a common issue in traditional RNNs. LSTM has become very popular in the field of deep learning, especially for tasks that require the processing of sequential data such as time series forecasting, natural language processing, and speech recognition.

\end{itemize}

\section{RRAM for AI}

In recent years, RRAM has been widely used in artificial neural networks. In order to achieve high inference accuracy, neural networks usually require a large number of parameters in which RRAM outperformed the traditional  structure of memory and computing points. An RRAM-based neural network integrates computing and storage closely, eliminating data transmission between the processor and memory, thus improving the overall system performance and saving most system energy consumption.

\begin{figure}
  \centering
    \includegraphics[width=0.8\linewidth]{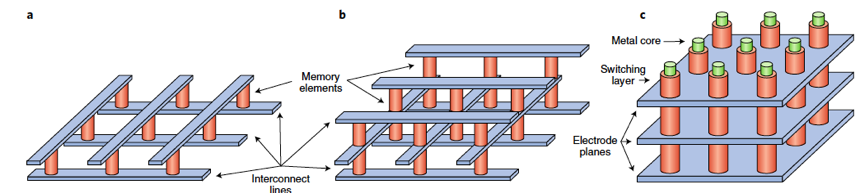}
  \caption{RRAM across-array architecture and scaling.25(a) RRAM across-array structure with a single memory layer. (b) Horizontal stacked 3Dacross-array structure. (c) Vertical 3D across-array structure. \cite{ielmini2018memory}}
  \label{im3}
\end{figure}

Metal oxide-based RRAM can act as the basic structure of a memristor which can then be constructed as an across-array structure to improve integration. The across-array structure of RRAM, as shown in Figure~\ref{im3}, is one of the bases of its fast calculations and responsible calculations. Employing a 3D structure can make resource-expensive tasks into a manageable size and provides substantial improvement to the speed and energy efficiency while running complex neural network model like multilayer perceptron (MLP) neural network, convolutional  neural  network  (CNN) and recursive neural networks (RNNs).

\begin{figure}
  \centering
    \includegraphics[width=0.8\linewidth]{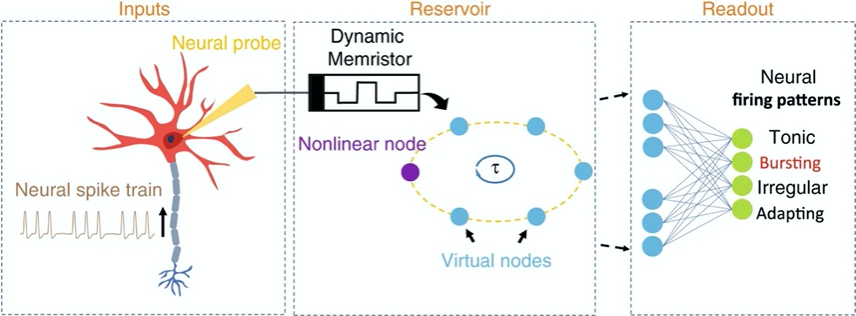}
  \caption{Schematic showing the concept of a memristor-based RC system for neural activity analysis.} 
  \label{im4}
\end{figure}

\begin{itemize}[leftmargin=*]
\item \textbf{Multilayer Perception (MLP)} Using RRAM across arrays to realize matrix-vector multiplication between the input information vector and weight matrix in parallel in one step can greatly reduce the energy consumption of the hardware neural network. Bayat \cite{bayat2018implementation} produced two 20×20 metal oxides RRAM (\ce{Pt}/\ce{Al2O3}/\ce{TiOx}/\ce{Ti}/\ce{Pt} memristor at each crosspoint) across-array to implement a multilayer perceptron (MLP) neural network. The MLP network has 16 inputs, 10 hidden layer neurons, and 4 outputs, enough to classify 4×4 pixel black and white patterns into 4 categories. Lately, Kim \cite{kim20214k} proposed a more complex array structure based on 64 × 64 passive crossbar circuit with \ce{Al2O3}/\ce{
TiOx} as its switching layer.

\item \textbf{Convolutional Neural Network (CNN)} The CNN based on memristor cross arrays mainly consists of the convolutional operation part and the fully connected layer part. The memristor array can store convolution kernels and complete the matrix-vector multiplication of input information and convolution kernels in one step, which greatly improves computational efficiency. Yao \cite{yao2020fully} implemented a five-layer convolutional neural network based on 2048 RRAM (TiN/TaOx/HfOx/TiN) array chips. The chip was two orders of magnitude more energy efficient than an image processor (GPU) in processing convolutional neural networks by comparison with conventional neural network computation.

\item \textbf{Recursive Neural Networks (RNNs)} Researchers have successfully used memristors for reservoir computing (RC) which is a concept developed from recursive neural networks (RNN). Figure~\ref{im4} shows the concept of a memristor-based RC system \cite{zhu2020memristor}. The spikes collected from firing neurons are used directly as inputs to an excitation memristor. The repository space is further extended with the concept of virtual nodes to help handle complex time inputs. A simple neural network is used as the reservoir's readout layer to produce the final output. This work makes it possible to realize efficient neural network signal analysis with high spatial and temporal accuracy which have a significant impact in the field of brain-like computing.

\end{itemize}

\section{Future Directions and Challenges}

The combination of metal oxides as oxide layers for RRAM and AI is a promising approach to improve the performance of RRAM devices and enable new applications in AI.  With the constraints of the von Neumann structure and the physical limits of semiconductor processing, computers are experiencing bottlenecks in their ability to process data. As a new computing principle, RRAM-based computing-in-memory technology can embed the computation capability into the storage unit thus increasing  computation  speed, reducing  system  power consumption, and opening up more application possibilities. This all gives RRAM an opportunity to boost AI. while AI algorithms can also be used to optimize the performance of RRAM devices. The combination of both can lead a rapid development in this area.

 For example in developing sense-storage-computing fusion technology. While the human nervous system senses the physical world in an analog but efficient way, researchers have been working on using hardware to simulate human perceptual systems. Sensors take signals such as light, pressure, and sound and convert them into electrical signals. The processor is used to analyze electrical signals and simulate human sensory systems. As a new type of memory, RRAM can be made of materials that are sensitive to peripheral signals and can be used as a sensor. While AI can also be used to help find these materials that are sensitive to peripheral signals and can be used as a sensor. At last, RRAM can combine micro-sensors with neural network computing to realize the sense-storage-computing fusion technology.

Although significant progress has been made in the combination of metal oxides for RRAM and AI, there are still some challenges that need to be addressed. For RRAM-powered AI, the stability and endurance of RRAM devices need to be improved to meet the requirements of practical applications. Moreover, the integration of RRAM devices and AI algorithms is still in the early stages, and further research is needed to optimize their performance.
Although across-array can greatly compress the chip volume and improve the computing efficiency and computing volume in integrating RRAM, the problem is also obvious and there are still unresolved problems: chip design and preparation processes are difficult, and every application needs to be customized. There is still a lot of room for development. For AI-assisted RRAM, though machine learning algorithms have been used for device performance improvement and some fabrication processes, few metal oxide materials have been tried to verify its widespread efficiency; meanwhile, more advanced ML algorithms than current basic ones can be tried to improve the final performance, like using graph neural network (GNN) to represent the material structure, trying gated recurrent unit (GRU) in handling temporal sequence of historic data to predict the current value and using deep learning method in the regression and classification works; the building of an open source dataset should also be considered to fast the development of AI-assisted RRAM performance improvement. Last but not least, AI can be tried in more steps in the fabrication process or even the whole fabrication process.

Although there are still challenges to be addressed, the integration of metal oxides for RRAM and AI is a promising research direction for future computing systems.

\bibliographystyle{abbrv}
\bibliography{ref}

\clearpage
\begin{landscape}
\begin{table}[]
\caption{Comparison between different works.}
\small
\begin{tabular}{|p{2cm}|p{2cm}|p{4cm}|p{4cm}|p{4cm}|p{4cm}|p{4cm}|}
\hline
Authors   & Oxide-layer   materials    & \multicolumn{1}{c|}{database}     & \multicolumn{1}{c|}{ML techniques}    & \multicolumn{1}{c|}{input / output}   & \multicolumn{1}{c|}{Evaluation   methods}                                                              \\ \hline
Zhang et al., 2020   & \ce{HfO2} (train), \ce{NiO}, \ce{TiO2}, \ce{SnO2}, \ce{VO2}, \ce{HfO2} and \ce{Ta2O5} (test).                                     & 1835 sets of phase-field simulations.                                                                                                 & compressed-sensing   based machine learning (performance prediction)                                                                                          & $V_{ij}$, $\sigma$, $k_{th}$ / $I_{on}/I_{off}$, $t_{switch}$                                                    & root-mean-squared-error (RMSE)                                                                        \\ \hline
Lin et al., 2021                        & HfO2                                     & 150000 measured I-V data points from real devices                                                                                                                        & MLP and LSTM model (compact modeling for RRAM)                                                         & sinusoidal and random walk signal-shaped voltage sequence, annealing temperature (T) / current value.                              & RMSE, R-2 score, MAE, and RAE                                                                   \\ \hline
Lin et al., 2022                        & HfO2                                     & One are actual measured RRAM I-V data. The other is ideal RRAM I-V data with changing set/reset. & 2 MLP (one for state variable W predicting, another for I predicting)                         & Discrete and continues state value W, voltage sequence V and sweeping speed r. / current value.            & RMSE and R-2.                                                                       \\ \hline
Hutchins et al., 2022                   & HfOx                                     &KMC-TCAD simulations/experimental data with duplicated and added gaussian noise                                                 & MLP (compact models for RRAM)                                                                                        & The voltage, state, and device parameters (such as device size, operating temperature,   etc.) / current value.                            & RMSE and R-2                                                                \\ \hline
Dongale et al.,   2022 & \ce{HfO2/x}, \ce{Ta2O5}, and \ce{TaOx} & dataset derived from 15,000 publications between 2007 and 2020.                & CART-based ML model and random forest algorithm (modeling the continuous features), and decision tree (DT) techniques (modeling categorical features). GB algorithm (Feature importance calculating)
 & TYM, SM, TE, TTE, BE, TBE, and TSL. / VSET,  VRESET,  EC,  RT, and MW (continuous); TSAD, TSUB, MRS, CM, and RM (categorical).  &   Pearson’s  r  and  Adj.  R2 (continuous  output); confusion matrix with accuracy and misclassification (categorical  output  features).                                                                                                                               \\ \hline
Santosh et al.,  2022 & \ce{CuO}, \ce{CuOx}, \ce{Cu2O}, \ce{CuxO}   & 935 data points from 55 research articles published between 2008-2022          & CART algorithm and random forest algorithm (modeling the continuous features);
DT algorithm (modeling categorical features ); RF, ANN, and LM algorithms (performance prediction).
        & TYM, SM, TE, TTE, BE, TBE, and   TSL. / VSET,  VRESET,  EC,  RT, and MW (continuous);
TSAD, TSUB, MRS, CM, and RM (categorical).
  & continuous  output: Pearson’s  r  and  Adj.  R2.
categorical  output  features:
confusion matrix with accuracy and misclassification
                                                 \\ \hline
\end{tabular}
\label{table2}
\end{table}
\end{landscape}


\end{document}